\documentclass[epjCONF]{svjour}
\usepackage{graphics}
\usepackage[varg]{txfonts} 
\usepackage[latin1]{inputenc}

\session-title{MESON2012 - 12th International Workshop on Meson Production, Properties, and Interaction}

\begin{document}
\title{Radial and angular-momentum Regge trajectories: a systematic approach}
\author{P. Masjuan\inst{1},\inst{2}\fnmsep\thanks{\email{masjuan@kph.uni-mainz.de}. 
Supported by MICINN of Spain (FPA2006-05294), CPAN (CSD2007- 00042), Junta de Andaluc\'ia (FQM 101, FQM 437, FQM 225, and FQM 022), by the Deutsche Forschungsgemeinschaft DFG through the Collaborative Research Center ``The Low-Energy Frontier of the Standard Model" (SFB 1044), by Polish Ministry of Science and Higher Education, grant N~N202~263438, and by National Science, grant DEC-2011/01/D/ST2/00772.}  
\and E.R.Arriola \inst{3} 
\and W. Broniowski \inst{4},\inst{5} }
\institute{Departamento de F\'isica Te\'orica y del Cosmos and CAFPE, Universidad de Granada, E-18071 Granada, Spain \and Institut f\"ur Kernphysik, Johannes Gutenberg-Universit\"at, D-55099 Mainz, Germany \and Departamento de F\'isica At\'omica, Molecular y Nuclear and Instituto Carlos I de F\'isica Te\'orica y Computacional, Universidad de Granada, E-18071 Granada, Spain \and The H. Niewodnicza\'nski Institute of Nuclear Physics, PL-31342 Krak\'ow, Poland \and Institute of Physics, Jan Kochanowski University, PL-25406 Kielce, Poland }

\abstract{
We present the analysis of Ref.~\cite{MAB} of the radial ($n$) and angular-momentum ($J$) Regge trajectories for 
all light-quark meson states listed in the Particle Data Tables. The parameters of the trajectories are obtained 
with linear regression, with weight of each resonance inversely proportional to its half-width squared, $(\Gamma/2)^2$. 
The joint analysis in the $(n,J,M^2)$ Regge plane indicates, at the 4.5 standard deviation level, that the slopes in $n$ 
are larger from the slopes in $J$. Thus no strict universality of slopes occurs in the light non-strange meson sector. 
We also extend our analysis to the kaon sector.
} 

\maketitle
%


In Ref. \cite{Anisovich:2000kxa} it was suggested that the light-quark meson states could be grouped into \textit{radial} linear Regge 
trajectories with the slope $\mu^2=1.25(15)$~GeV$^2$, where the error was estimated as the spread of the values for each meson-family considered ($\rho,\pi,\eta,a,f$).
In Ref.~\cite{Afonin:2006vi} a joined formula assuming universality of slopes was proposed, $M^2(n,J)=b+a( n + J)$, with $a=1.14$~GeV$^2$.
In Ref.~\cite{MAB} we reanalyzed the radial and angular-momentum Regge trajectories with the updated list of the light unflavored 
mesons from the PDG \cite{PDG}. For the fits and error estimates we have used the \textit{half-widht rule} \cite{MAB,RuizArriola:2010fj}, i.e, the half-widht squared as a weight for each resonance.

\begin{figure}
\centering
\resizebox{0.738\columnwidth}{!}{%
\includegraphics{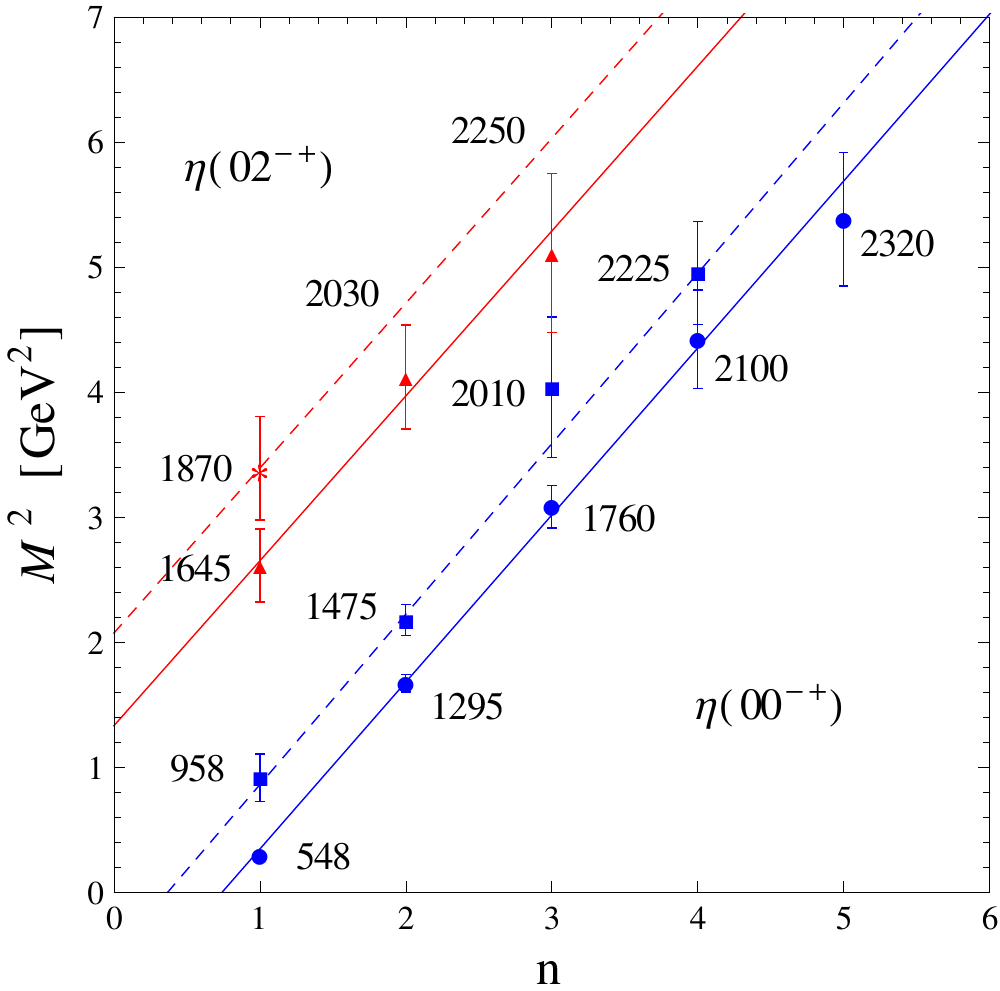}
\includegraphics{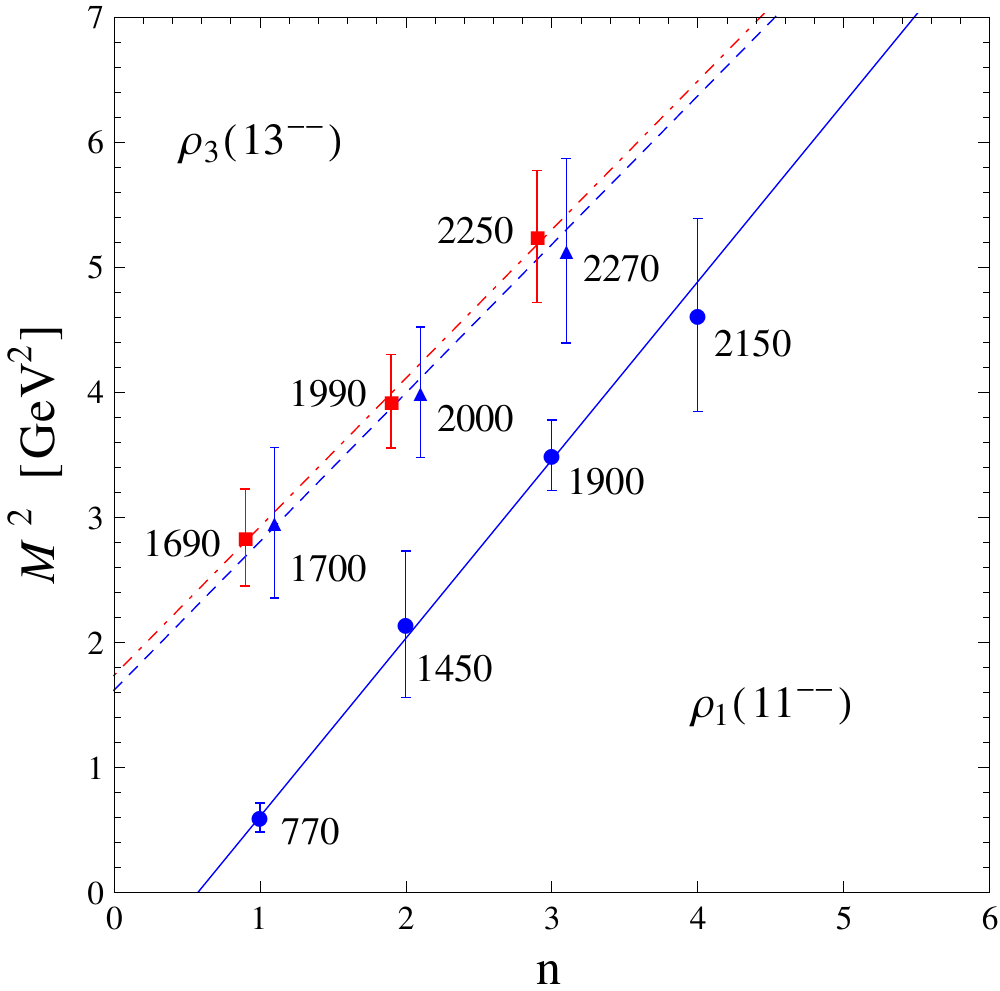}}
\resizebox{0.738\columnwidth}{!}{
\includegraphics{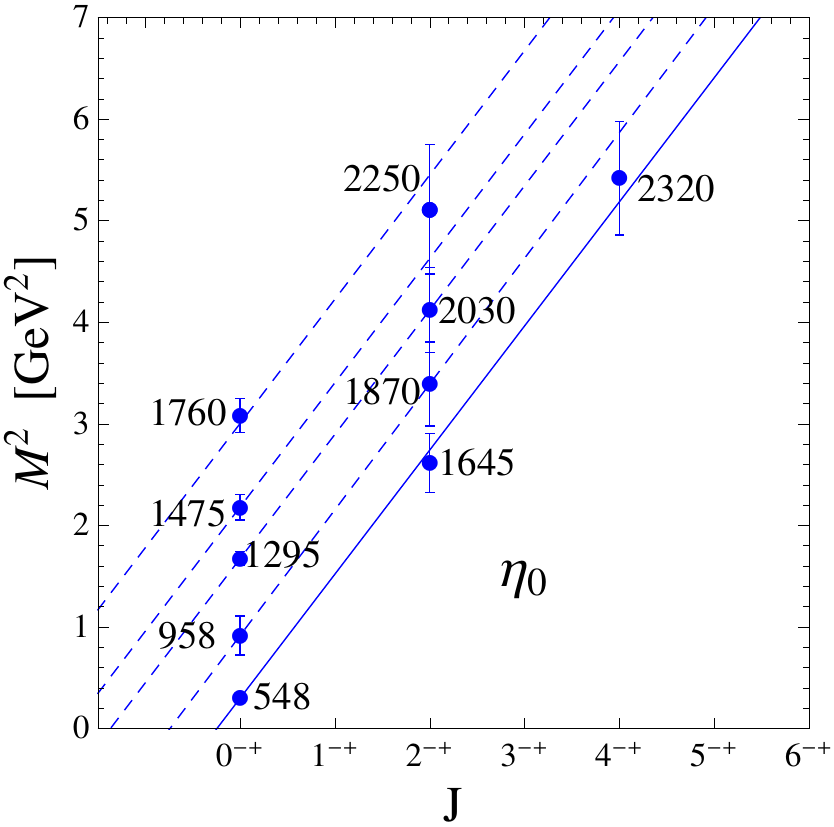}
\includegraphics{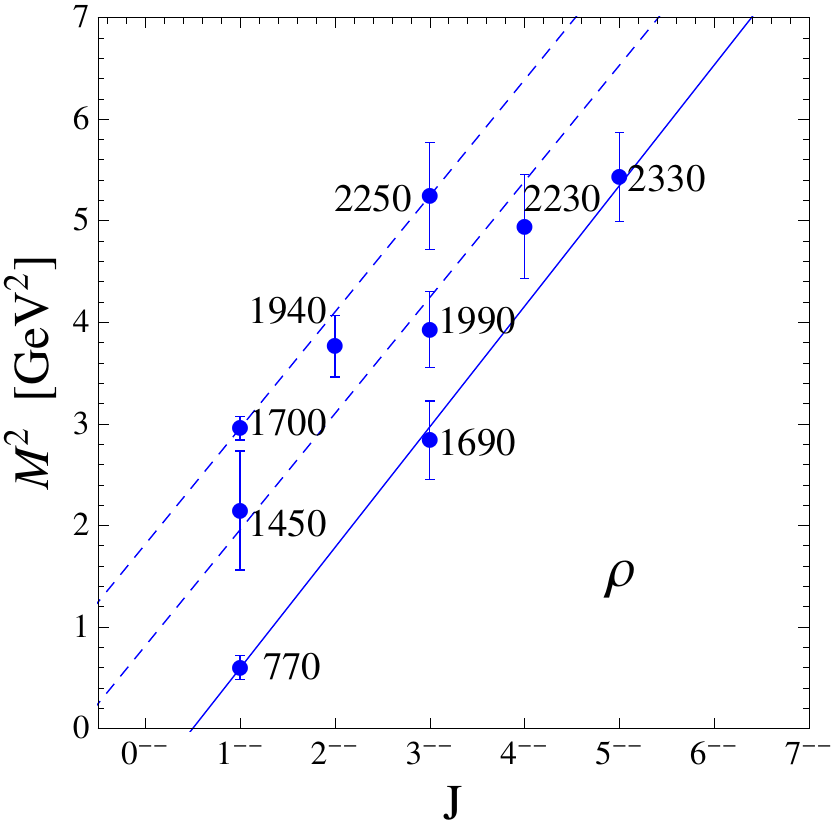}}
\caption{The $(n,M^2)$ and $(J,M^2)$ plots for the $\eta$ and $\rho$ meson families.}
\label{ntraj}       
\end{figure}

\begin{figure}
\centering
\resizebox{0.6\columnwidth}{!}{%
\includegraphics{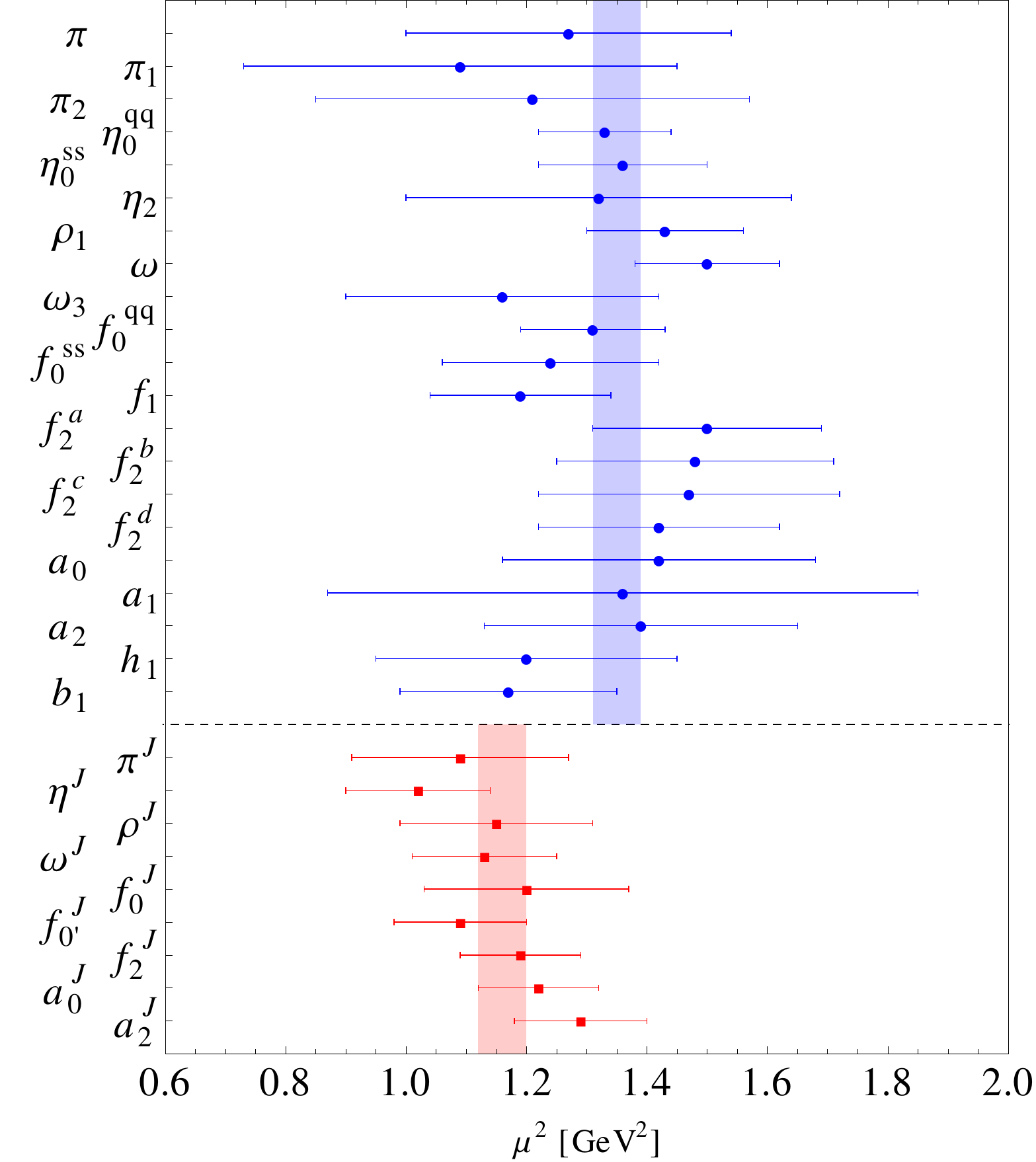} }
\caption{$(n,M^2)$ and $(J,M^2)$ slopes for the meson families considered in Ref.~\cite{MAB}.}
\label{Compilation}       
\end{figure}

The fit to all the light unflavored meson families with linear trajectories using the half-width rule yields 
$\mu^2=1.35(4)$~GeV$^2$ as the weighted averaged result for the slope of radial trajectories, and $\beta^2=1.16(4)$~GeV$^2$ as the weighted average for the slope of the angular-momentum trajectories (the bands in Fig.~\ref{Compilation}). Fig.~\ref{ntraj} exemplifies the results for the $\eta$ 
and $\rho$ families. In Fig.~\ref{Compilation}, we collect the slopes from both radial and angular-momentum trajectories from all the families considered.

We also considered a joint fit with the formula $M^2_X(n,J)=M^2_X(0,0)+ n \mu^2 + J \beta^2$,
with the result $M^2_X(n,J)=(-1.25(4)+ 1.38(4) n  + 1.12(4)J)$~GeV$^2$, which means a difference between the radial and the angular-momentum slopes at a statistically significant level of 4.5 standard deviations.

\begin{figure}
\centering
\resizebox{0.9\columnwidth}{!}{%
\includegraphics{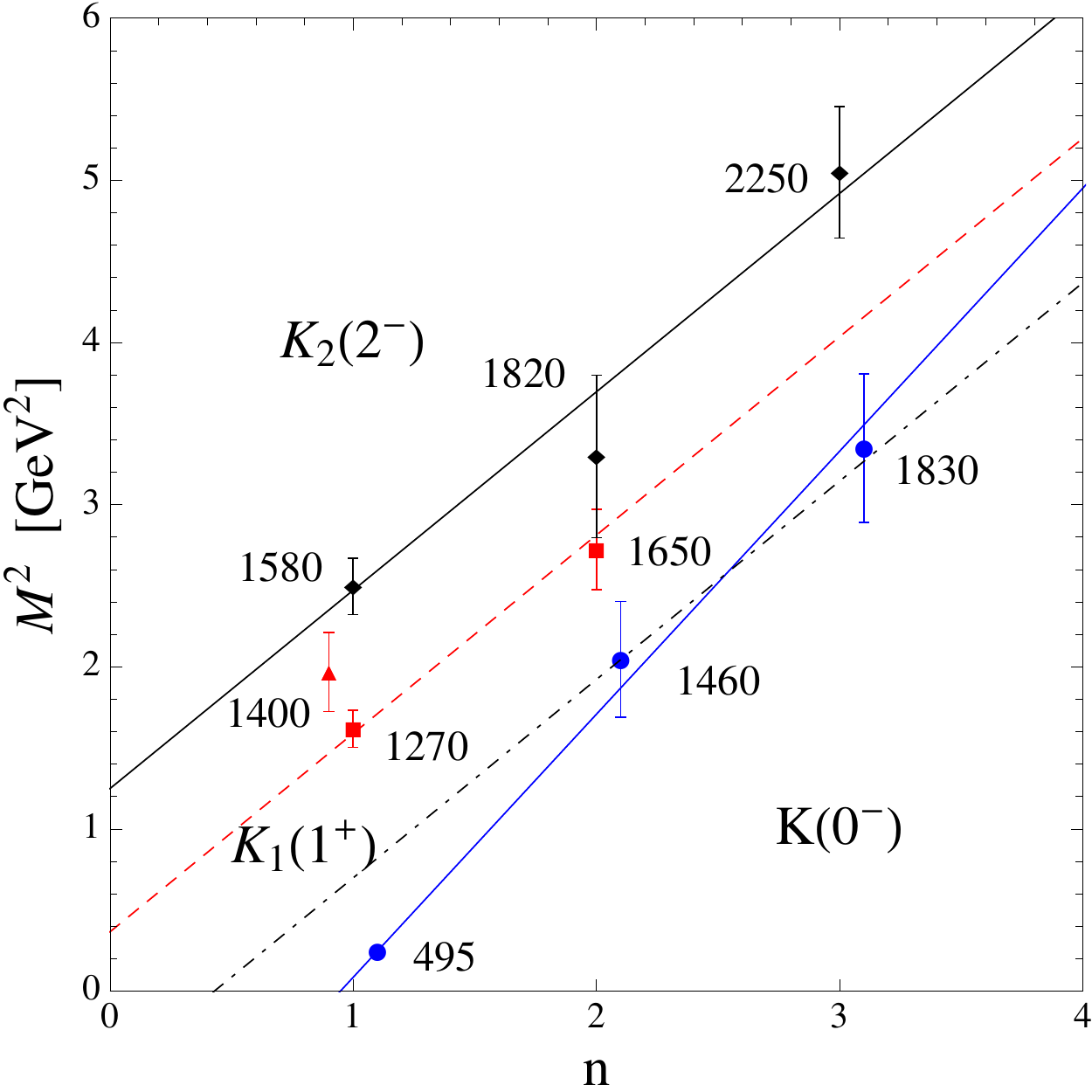}
\includegraphics{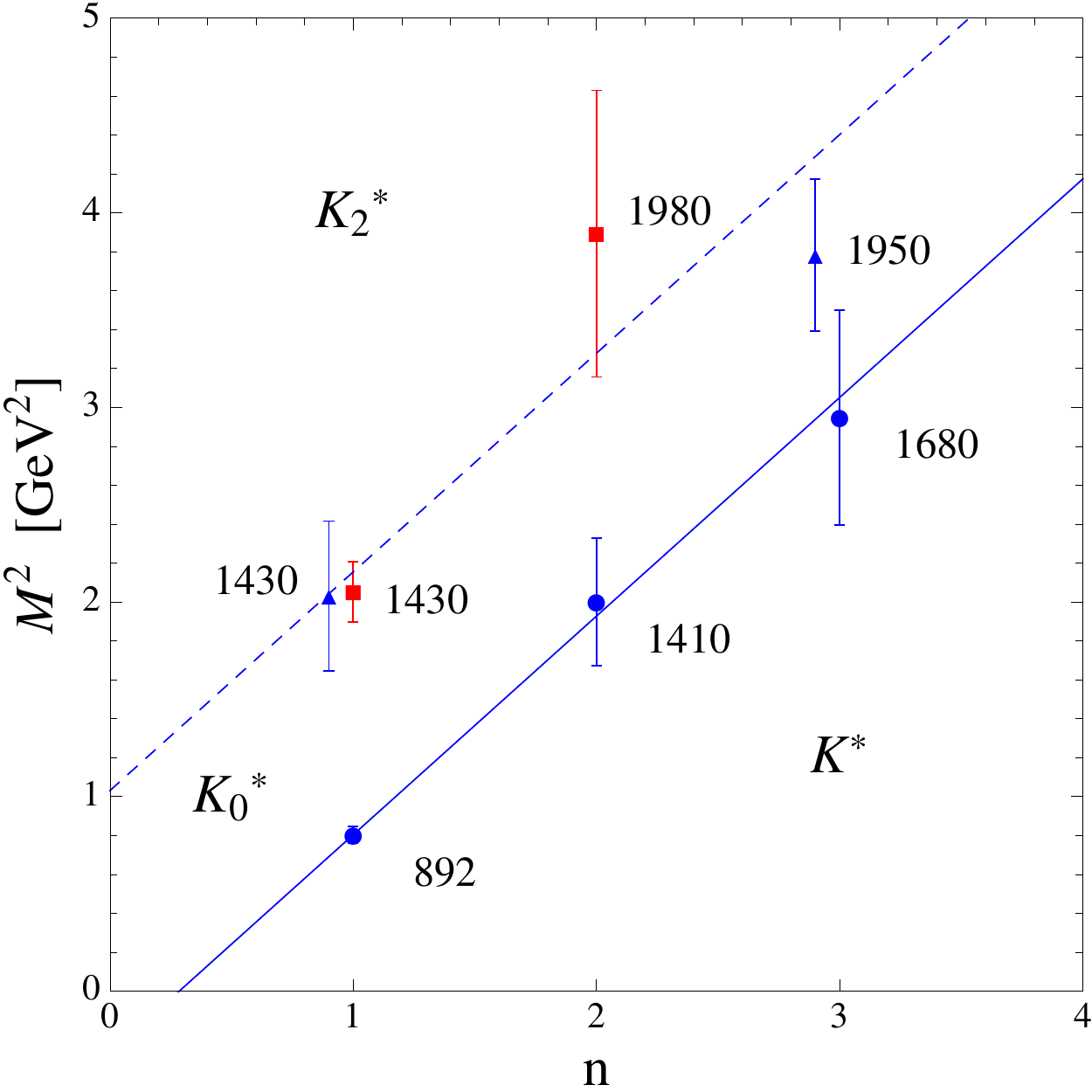}}
\resizebox{0.9\columnwidth}{!}{
\includegraphics{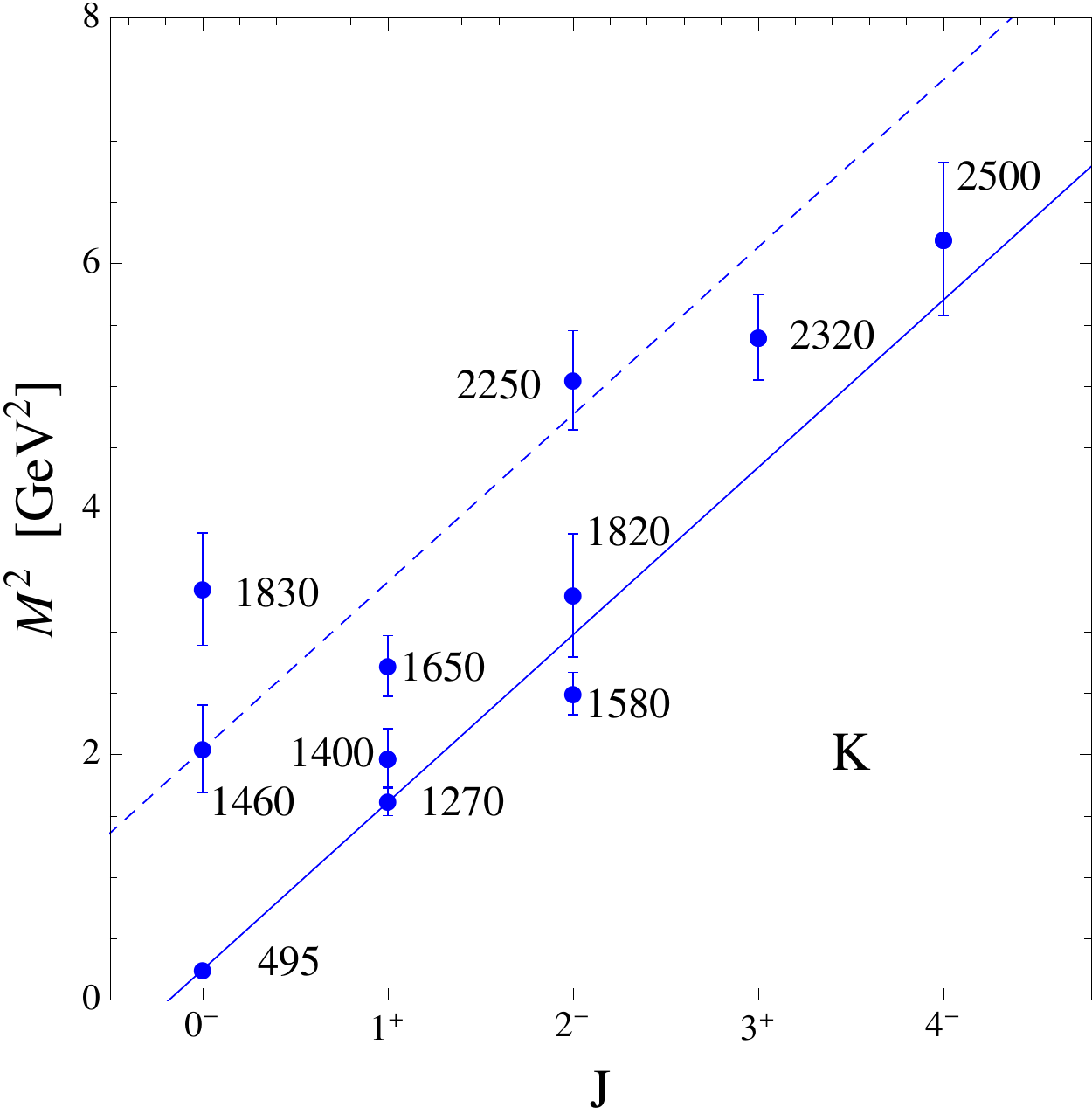}
\includegraphics{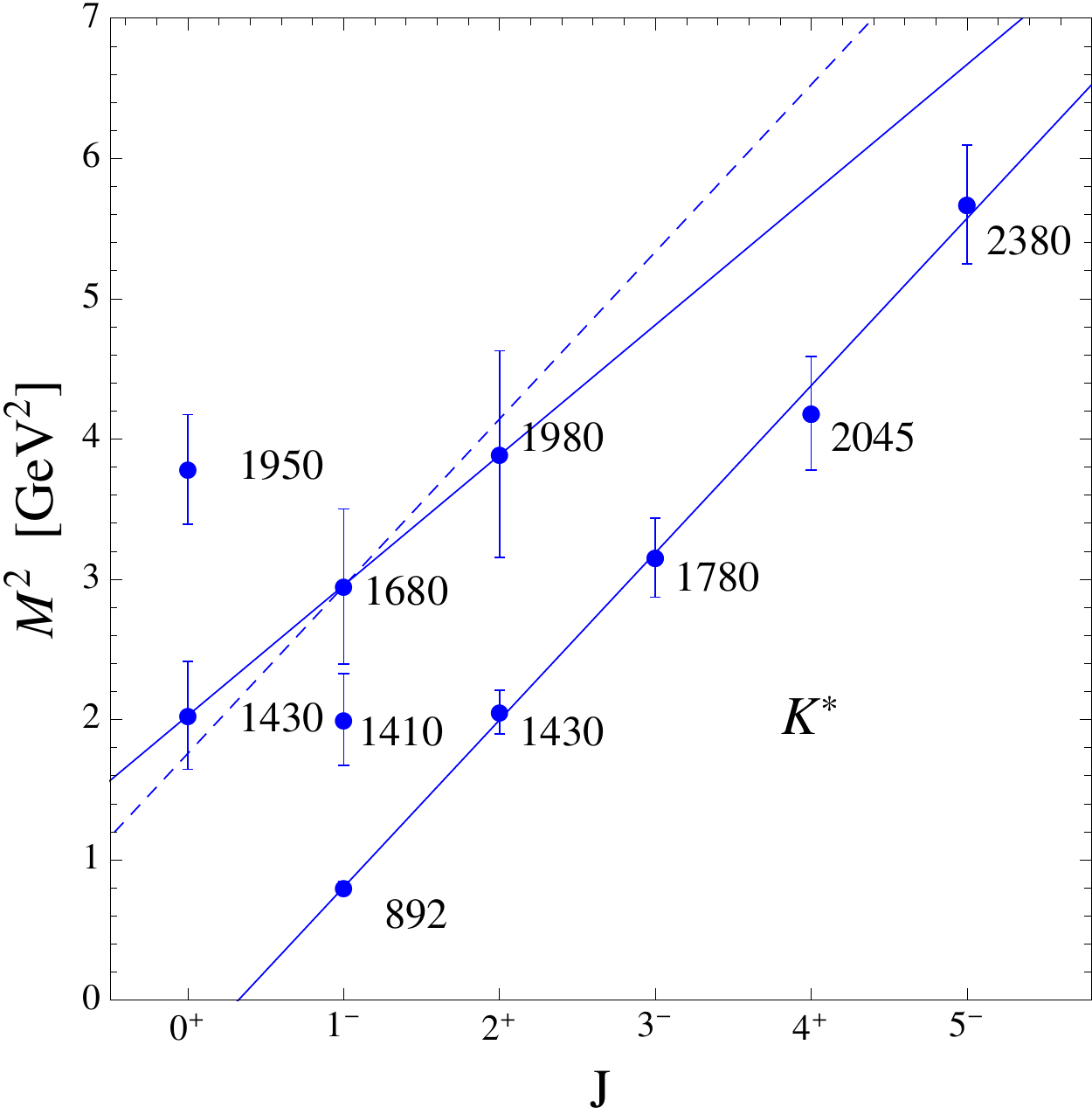}}
\caption{$(n,M^2)$ and $(J,M^2)$ slopes for the kaon sector. The error bars follow from the half-width rule.}
\label{Kaonplots}       
\end{figure}

As an extension of Ref.~\cite{MAB}, we present in Fig.~\ref{Kaonplots} a study of both radial and angular-momentum trajectories for the kaon sector. The radial fit yields $\mu^2_K=1.22(21)$~GeV$^2$ and $1.12(21)$~GeV$^2$ for $K$ and $K^*$, respectively, while the angular-momentum fit returns $\beta^2_K=1.36(6)$~GeV$^2$ and 
$1.19(7)$~GeV$^2$ for $K$ and $K^*$, respectively.  
Only trajectories containing more than three states are considered. 

\begin{thebibliography}{4}

\bibitem{MAB}
 P.~Masjuan, E.~R.~Arriola and W.~Broniowski,
  Phys.\ Rev.\ D {\bf 85} (2012) 094006.
\bibitem{Anisovich:2000kxa}
  A.~V.~Anisovich, V.~V.~Anisovich and A.~V.~Sarantsev,
  Phys.\ Rev.\ D {\bf 62} (2000) 051502.
\bibitem{Afonin:2006vi}
  S.~S.~Afonin,
  Phys.\ Lett.\ B {\bf 639} (2006) 258.
\bibitem{PDG}
  K.~Nakamura {\it et al.}  [Particle Data Group Collaboration],
  J.\ Phys.\ G G {\bf 37} (2010) 075021.

\bibitem{RuizArriola:2010fj}
  E.~Ruiz Arriola and W.~Broniowski,
  Phys.\ Rev.\ D {\bf 81} (2010) 054009;
    E.~Ruiz~Arriola and W.~Broniowski,
  (Bled Workshops in Physics. Vol. 12 No. 1), arXiv:1110.2863;
  P.~Masjuan, S.~Peris and J.~J.~Sanz-Cillero,
  Phys.\ Rev.\ D {\bf 78} (2008) 074028;
  P.~Masjuan,
  arXiv:1206.2549.
   
\end{thebibliography}
\end{document}